ArXiv 0910# Achievements and challenges of nanostructured Titania Eco-materials derived from Sol-Gel Processing

by

Wilfried Wunderlich[*1], Rino R. Mukti[2], Suminto Winardi[3],
Krishnankutty-Nair. P. Kumar[4], and Tatsuya Okubo[5](Received on Sept. 29, 2009 and accepted on XXX XX, 2009)**Abstract**
Nanostructured titania derived from the sol-gel processing requires a detailed study to clarify state-of-art of every process step starting from colloidal sol, peptization, gelation to the final sintering. The applied parameters such as peptization temperature and time in the current investigation of sol during the early stage of processing were characterized by dynamic light scattering (DLS) and compared to other parameters found in the literature. Upon gelation, the resulting anatase was calcined to study the phase transformation to rutile as well as to discriminate the effect of peptized and unpeptized in which the characterizations were carried out by using XRD, Nitrogen adsorption-desorption isotherm and HRTEM. It shows that peptizing the titania sol at 80 °C for at least 12 h resulted in a small and uniform anatase crystallite with size of 4.6 nm.

*Keywords:* Sol-gel ceramics processing, nano-technology, renewable energy from photocatalyst## 1. Introduction

After the Austrian chemist Richard Zsigmondy [1] studied colloids and received the Nobel Prize in 1925, this knowledge was applied for producing paint, crème, toothpaste and medicine. Colloid processing became again popular since the late 1980-ies after the sol-gel technique was developed, which allows processing of nano-sized titania. Nano-titania has many applications due to its superior photocatalytic [2, 3], photovoltaic, electro-chromic and sensor properties, as summarized in a long review paper [3]. Hence, nano-titania is considered as an environmentally benign material, which has two functions, namely as a part of a photovoltaic or photocatalytic energy conversion system as well as environment protector by self-cleaning photocatalyst properties [4]. Anatase is the desirable phase, because of its higher catalytic activity due to higher Fermi level compared to the stable phase rutile and the less recombination rate photo-induced electron-hole pairs [3, 5]. The anatase to rutile phase transition can be utilized for compaction during calcination or sintering [6, 7] even for structural ceramics, in spite of the volume decrease of 8% during anatase to rutile transformation [8].

The common process for obtaining nano-ceramics [3, 4, 6-11] is the peptized sol-gel method as summarized in fig. 1. The stabilized sol can be either used directly or after peptization usually performed with nitric acid. The

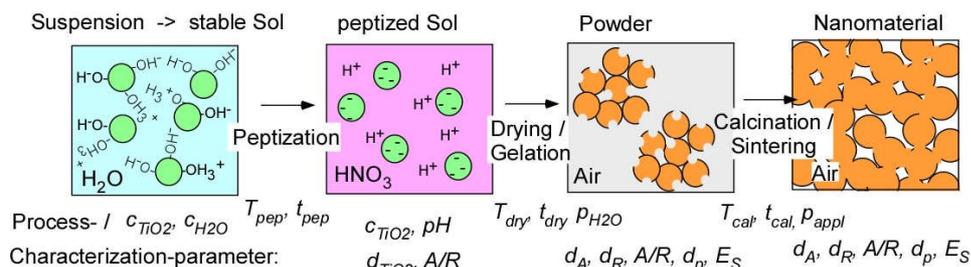

Fig. 1. Process parameter $c$ = concentration, $T$ = temperature, $t$ = time, $p$ = pressure are considered in this study for obtaining nano-sized anatase of rutile ceramics by sol-gel technology. The sol, powder and nano-material are characterized by the parameter $d$ = particle radius, A/R = anatase to rutile ratio $d_p$ = pore size, $E_S$ = surface energy.

[*1] Associate Professor, Department of Material Sciences, Tokai University, Hiratsuka-shi Kanagawa 259-1292, Japan
2 Doctor Student, Dept. Chemical System Eng., The University of Tokyo, Bunkyo-ku, Tokyo 113-8656, Japan
3 Graduate Student, Dept. Chemical System Eng., The University of Tokyo, Bunkyo-ku, Tokyo 113-8656, Japan
4 Professor, Dept. Material Science and Engineering, The University of Texas at Dallas, TX 75080-3021, USA
5 Professor, Dept. Chemical System Eng., The University of Tokyo, Bunkyo-ku, Tokyo 113-8656, Japan—1—

W. WUNDERLICH, R. MUKTI, S. WINARDI, K.N.P. NAIR, T. OKUBO

goal of this study is to show, how the process parameters ($c$=concentration, $T$= temperature, $t$=time, $p$=pressure) influence the microstruture of the nano-material characterized by the parameter $d$ = particle radius, A/R = anatase to rutile ratio, $d_p$ = pore size, $E_S$ = surface energy. Peptization acts as electro-static separation in order to break-down agglomerates into nano-particles. During drying the removal of the liquid can agglomerate particles again due to capillarity force. When drying is performed at temperatures not to high above room temperature, the surface of the particles are in an unrelaxed state as sketched in fig 1c. During powder calcination or sintering of pressed samples usually at temperatures around 300 to 600°C the particles are packed together by diffusion. At this stage the transformation into the stable rutile phase takes place, unless microstructural features suppress this transformation. The powder behavior strongly depends on its processing history and it is one of the goals of this research project to clarify this, as well as their physical properties dependence. Experiments [3] and calculation [10-13] showed that particle size, surface energy and pore diameter have a large influence on the band-gap and hence the performance of the photocatalysts. The goal of this paper is to clarify the influence of the parameters on the processing of nano-titania. Further improvement can be achieved by doping [3, 14-17] which is, however, beyond the scope of this paper.

The transformation behavior of stabilized sols into titania powder and their kinetics have been studied [3, 5, 18], especially the transformation behavior of anatase to rutile [3, 19-22]. As mentioned above, anatase is the desired phase for achieving good photocatalytic properties and one goal of processing is to stabilize anatase even after compaction. It has become common to use as many as analysis techniques as possible to characterize the sol, the calcined powder as well as the ceramics after sintering, but the later one is beyond the scope of this paper. With transmission electron microscopy (TEM) particle size, shape and morphology can be directly imaged [3, 12-17], and with electron diffraction also anatase and rutile phase can be distinguished [15]. The gas adsorption method has become the most widely used standard procedure for the determination of the pore size and surface area of porous materials and powders [3, 16, 17, 23].

This paper describes preliminary results of a new sol preparation technique and the dependence of microstructure on the process parameters, first for peptization, then for calcination. The properties are characterized by XRD, TEM and $N_2$ adsorption-desorption measurements.

**Experimental**

In this overview only a short summary of the experimental procedure is presented, for details see [24]. The particulate hydrosol of anatase was prepared by initially pre-hydrolyzing titanium isopropoxide in an alcoholic solution (isopropyl alcohol) of 0.55 M with an alcoholic solution containing water and isopropyl alcohol at room temperature (RT) under vigorous stirring. The precipitation reaction results in white titanium oxyhydroxide. The precipitates were filtered and washed with water to remove the excess of alcohol. In the next step the precipitates were re-dispersed in 183 ml water. Both of these washing steps are obviously important for achieving the superior properties described later. This sol is stable for at least four months as checked and is termed as unpeptized sol. It is used for the following sol peptization process.

Direct evaporation of this sol at 40 °C gave gel and is termed as unpeptized sample. In order to study the effect of peptization on nanocrystalline titania, the unpeptized sol was heated at 80 °C under reflux with vigorous stirring for 12 h, 24 h and 33 h. The peptization was performed under constant pH=1.5 by adding nitric acid with $H^+/Ti^{4+}$ ratio of 0.5 to this sol, which finally changed its color from whitish to light-blue. Subsequently, evaporation at 40°C was applied to each such treated sample in order to obtain the final gel. The nano-powder of peptized titania samples were calcined for 8 h with three different temperatures at 400, 500, 600°C in a muffle furnace exposed to atmospheric air. The phase transformation of anatase to rutile, at least partly, occurred in this temperature range.

The mean particle size and its distribution in the sol were measured using dynamic light scattering (DLS). X-ray diffraction patterns were recorded using Bruker-AX; M0X3X-HF system with CuK$\alpha$ radiation. The diffractograms were performed using a step scan rate of 4 °/min.

Nitrogen adsorption-desorption isotherms were measured using a sorption apparatus (Autosorb-1 from Quantachrome Instruments). All samples were degassed at 110 °C prior to the measurement. The measurement was conducted using $P/P_o$ within the $10^{-3}$ to 1 region. The results of peptized titania showed sroption isotherm of Type I, microporous characteristic and after the calcination isotherm altered into Type IV with the presence of hysteresis loops indicating to the capillary condensation, typical for mesoporous characteristic. Surface area values and pore size distributions were calculated based on BET method [23, 24]. The crystalline fringes of anatase specimens could be observed using a field-emission high-resolution transmission electron microscope (FE-HRTEM) from Hitachi HF 2200 operated at an accelerating voltage of 200 kV. For the sample preparation, powder samples were mixed with ethanol, dipped on a carbon-coated Cu-grid and dried under air, while in the case of sols, the TEM-grid was directly dipped into the sol and dried at room temperature.

**Results and Discussion**

According to literature data [3, 5, 18], there is a strong dependence of the sol particle size on the peptization time. The breaking of the agglomerates during peptization takes a certain time and the saturation of the smallest particle size will be achieved after a certain time period [18]. Recently developed methods, like dynamic light scattering (DLS) and laser diffraction (LD) and near-infrared (NIR) light transmission allow the direct study of sol properties, as described in detail elsewhere [5, 18]. Fig. 2 was plotted with peptization time data from [5, 18] until the point of saturation is reached. Fig. 2 shows that there is a strong dependence on the peptization temperature and the Ti-concentration in the sol. As our peptization temperature was °C (353 K) for 12, 24 and 33 h, therefore we confirm that the saturation state towards small particle size should have been reached. The particle size distribution achieved in our peptization study was combined together with the literature data [5, 18] (Fig. 3) in order to understand role of temperature during peptization as the first goal of this paper. The particle size distribution was measured as the 50% value of the size distribution obtained from DLS measurements. The particle size after peptization achieved in our study is plotted in fig. 3 together with literature data [5, 18]. The particle size was measured as the 50% value of the size distribution obtained from DLS measurements. The data in reference [18] were measured under different Ti-concentration in [g/cm3] under constant pH=1.1, or in other words, the [H+]/[Ti] ratio is another key parameter besides peptization time and peptization temperature. The smallest particles can be achieved by lowering the Ti-concentration and the peptization temperature. As we measured only one data point of peptization temperature, the same slope as the data in [5] is assumed and indicated as an eye-guide line. The reason why the particle size in our study is nearly one order of magnitude smaller than in [5, 18] is attributed to relatively small concentration of Ti and high peptization temperature as well as the washing procedure as described above that certainly affects the the rheological behavior of our suspension, namely the viscosity.

Thereafter the specimens were dried at 40°C into flake form of gel and the powder form of sample was used for XRD-characterization. It is well known, that nano-sized particles can have different surface energy, which is usually higher than bulk values [3, 10, 11]. This has a large influence on the transformation behavior and it might even be necessary to revise bulk phase diagrams [11]. Hence, the second goal of this study is to check the transformation behavior of anatase to rutile on powders achieved under the described conditions.

The XRD patterns of peptized and unpeptized Titania are shown in Fig. 4. Both peptized and unpeptized Titania sol were aged at 80 °C for 33 h, followed by the gelation process under evaporation at 40 °C. The resolved powder was calcined at 400, 500 and 600 °C for 8h. The XRD pattern showed pure anatase, but at 400 °C already rutile peaks are observed, which intensities increase with increasing calcinations temperature. From the XRD peaks the crystallie size was calculated using the method using Scherrer's formula [20] and the fraction of rutile was determined from the area integral. The crystal size of anatase was determined as 4.6± 0.38 nm considering the saturation state of peptizing at 80 °C for 12, 24 and 33 h.   The data in fig. 5 show a part of the kinetic phase diagram. The anatase phase on the left side transforms to rutile phase on the right side in the mentioned volume ratios for different calcination temperatures after different time periods. Our data are summarized together with literature data showing also the calcinations time dependence for bulk material [19] and for 40nm sized sol-gel derived Titania [20]. In both cases the slope in this logarithmic plot is equal, so we assumed the same for our data, which were measured only for 8 h calcination time. The result shows that bulk material requires a higher temperature or longer annealing time in order to achieve the anatase to rutile formation. In 40 nm sized powder, however, the required temperature is much lower, while in the case of our 4.6 nm powder the desired anatase phase is stabilized at least until 400°C. For 40 nm sized nano-particles the anatase-to-rutile transformation occurs already at 400-500°C, because the short diffusion paths speeds up the reaction. For the 4.6 nm sized powder the beginning of the reaction is at similar time periods, but the retardation in the late transformation state is even more pronounced yielding to still remaining 15% untransformed anatase at a sintering temperature of 600°C. These results confirm the nano-size effect [10, 11] for particles smaller than 10 nm. Due to their metastable surface structure the transformation enthalpy is reduced, which even changes phase transformation temperatures. How this is correlated to the surface energy requires further investigations [24].

The third goal of this paper is to characterize the microstructure by TEM investigations. The sol was directly dropped on a Cu-grid used for TEM and dried at room temperature. The liquid evaporates and leaves an unique finger shaped pattern on the carbon coated grid as shown in fig. 6 (a). The reason for this self-structuring nano-pattern is its high viscosity, low interface energy to the substrate, high surface energy to air and low vapor pressure of the sol. The pattern can be used as template for further nano-material processing. Only on rims of the carbon net the sol has no other place to retract and forms thick bulges (fig. 6 a). The HRTEM micrograph in fig 6 b confirmed the lattice fringes of 4 nm sized, round-shaped anatase particles and 10-20 nm rutile particles with high aspect ratio. An unsolved question is whether the rutile particles are already present in the sol or form just after gelation.

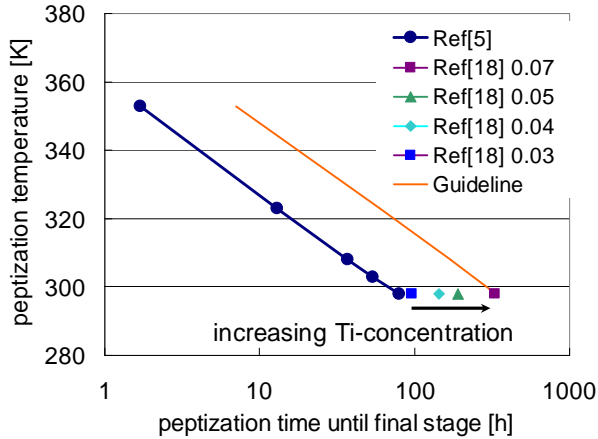

Fig. 2. Peptization time $t_{pep}$ for a certain temperature $T_{pep}$ for breaking down the agglomerates until the final stage of smallest particle size is reached. Data are shown from literature [5] and [10], in which the dependence on Ti concentration was also studied. The conditions for peptization in our study were 353 K at 12h, 24h, and 33h.

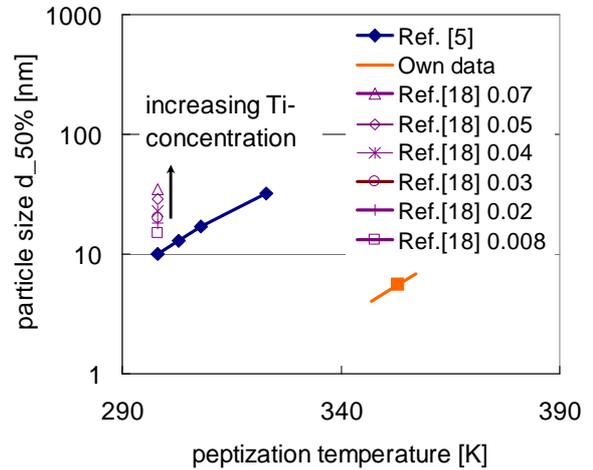

Fig. 3. Logarithmic plot of the sol particle size $d_{50\%}$ in nanometer after peptization under a certain temperature until final state is reached. Literature data [5.18], in which also the dependence on Ti concentration in [g/cm³] was studied, are shown together with our data.

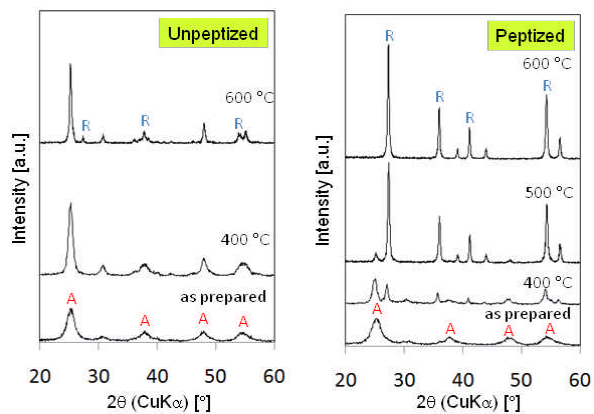
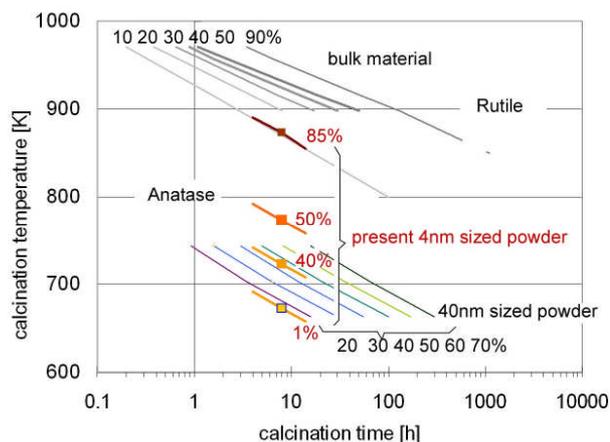

Fig. 4. XRD data of titania powder derived from a) unpeptized sol, b) peptized sol and calcined at different temperatures as mentioned. A= Anatase, R= Rutile.

Fig. 5. Logarithmic plot of calcination time versus calcinations temperature shows the transformation behavior of anatase (left) to rutile (right) for our present data on 4.6 nm sized powder compared to literature data on bulk material [19], and 40nm sized titania powder [20]. The numbers in % indicate the amount of rutile transformed from anatase.

Finally, fig. 7 shows the nitrogen absorption curves for unpeptized and peptized material. Specimens displayed with black lines were prepared under ambient conditions, and those with red liness were prepared after drying the sol at 80°C. Data points on as received powder are displayed with round symbols; those of calcined specimens (600°C for 8 h) are shown with triangular shape. The interpretation shown in purole color leads to type I data [23] characteristic for micropores with size less than 2nm in most cases. The point of slope change is usually used as a measure for the pore size [23]. After calcination the pattern changes to type IV characteristic for mesopores in the size of 2-50 nm.

Preliminary results [24] on the estimation of surface area by the Brunauer-Emmett-Teller (BET) method, showed rather high values of 208 $m^2/g$ for both cases, peptized and unpeptized compared to other materials. Hence these powders are suitable to produce materials, which can be used for photocatalysts or in dye sensitized solar cells. size of 2-50 nm. The external surface area was also 208 $m^2/g$ and further studies [24] showed that the history of the peptization influences the powder surface properties even after calcinations.

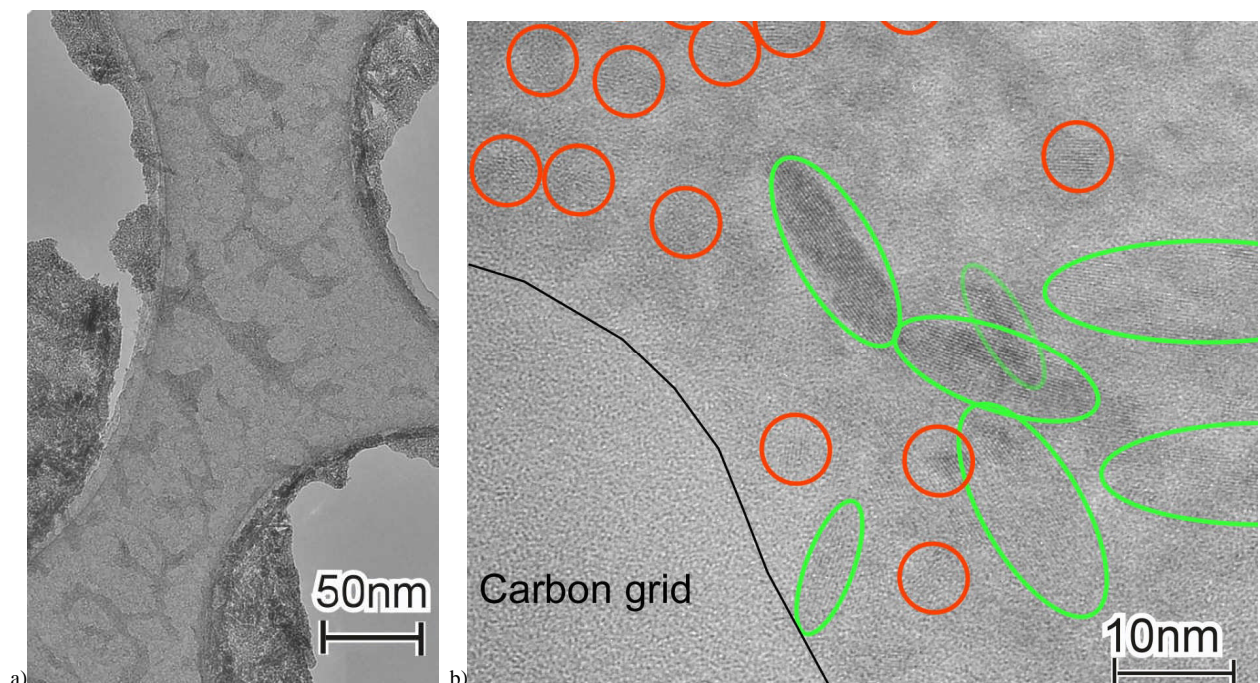

Fig. 6. TEM micrographs of gel dried on directly on the carbon grid used for TEM-observations; a) low magnification of the finger-shaped pattern on the carbon grid, b) high resolution. The dark elliptical particles consist of rutile, the others of anatase.

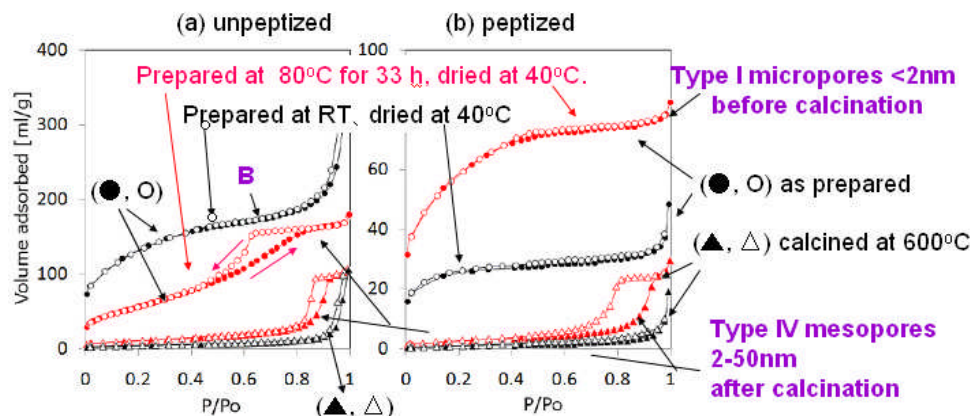

Fig. 7. Nitrogen adsorption/desorption plots of a) unpeptized, b) peptized gel dried at 40°C. The dark curves show specimens prepared at room temperature, the red curves show specimens peptized at 80°C. Round shaped data points are as prepared Triangular shaped data points are derived on calcined specimen at 600°C.

**Conclusion**

In this study the properties of a sol prepared with a new peptization method were compared with literature data and show three outstanding properties. The peptization time until the smallest particle size is reached, and this particle size, both decrease with temperature and concentration. By using titanium isoprop-oxide after a special washing procedure and peptization at 80°C we achieved a particle size 4.6 nm, much smaller than comparable literature data. In the subsequent calcination step at 600°C for 8 h these 4.6 nm sized particles shows a retardation of the anatase-to-rutile transformation and a volume amount of 15% anatase remains, which is much larger amount than at usual processed nano-sized powder. The third outstanding property is the gelation behavior of this sol. A unique finger-shape pattern with 40nm spacing is observed on carbon surfaces, which can be used as template for nano-scale processing.

**Acknowledgement**

This paper is written for the occasion of an invited talk at the ICFAM 2009 conference in Tiruvanatapuram, India, which is gratefully acknowledged.